\documentclass[aps,prl,showpacs,superscriptaddress,twocolumn]{revtex4}
\usepackage{amsmath}
\usepackage{graphicx}
\usepackage{psfrag}

\begin{document}

\title{Spintronic single qubit gate based on a quantum ring with spin-orbit
interaction}
\author{P\'{e}ter F\"{o}ldi}
\affiliation{Department of Theoretical Physics, University of Szeged, Tisza Lajos k%
\"{o}r\'{u}t 84, H-6720 Szeged, Hungary}
\affiliation{Departement Fysica, Universiteit Antwerpen (Campus Drie Eiken),
Universiteitsplein 1, B-2610 Antwerpen, Belgium}
\author{Bal\'{a}zs Moln\'{a}r}
\affiliation{Departement Fysica, Universiteit Antwerpen (Campus Drie Eiken),
Universiteitsplein 1, B-2610 Antwerpen, Belgium}
\author{Mih\'{a}ly G. Benedict}
\affiliation{Department of Theoretical Physics, University of Szeged, Tisza Lajos k%
\"{o}r\'{u}t 84, H-6720 Szeged, Hungary}
\affiliation{Departement Fysica, Universiteit Antwerpen (Campus Drie Eiken),
Universiteitsplein 1, B-2610 Antwerpen, Belgium}
\author{F. M. Peeters}
\affiliation{Departement Fysica, Universiteit Antwerpen (Campus Drie Eiken),
Universiteitsplein 1, B-2610 Antwerpen, Belgium}

\begin{abstract}
In a quantum ring connected with two external leads the spin properties of
an incoming electron are modified by the spin-orbit interaction resulting
in a transformation of the qubit state carried by the spin.
The ring acts as a one qubit spintronic quantum gate
whose properties can be varied by tuning the Rashba parameter of
the spin-orbit interaction, by changing the relative position of
the junctions, as well as by the size of the ring.
We show that a large class of unitary transformations can be attained
with already one ring -- or a few rings in
series -- including the important cases of the Z, X, and Hadamard gates.
By choosing appropriate parameters the spin transformations
can be made unitary, which corresponds to lossless gates.
\end{abstract}

\maketitle

The electron spin degree of freedom is one of the prospective carriers \cite%
{Aw,DD90} of qubits, the fundamental units in quantum information
processing. In order to implement quantum operations on electron spins,
appropriate gates are necessary that operate on this type of qubits. We note
that in the present context the word `gate' stands for an elementary logical
operation \cite{NC00}. In this paper we show that a one dimensional ring 
\cite{ALG93} connected with two external leads made of a semiconductor structure
\cite{VKDM04}, 
such as InGaAs in which Rashba-type \cite{Ras} spin-orbit interaction is the
dominant spin-flipping mechanism, can render such a gate. Conductance  
properties of this kind of rings have been discussed earlier in the case of
diametrically connected leads  \cite{NATE97,NMT99,MPV}. 

By taking here a  new point of view, we focus explicitly
on the spin transformation characteristics 
of this device, and show that those can be appropriately controlled by 
varying its geometrical and physical parameters in the experimentally 
feasible range \cite{NATE97,NMT99}.  
We shall determine the effects of changing the radius and the relative positions
of the junctions, as well as the influence of varying 
the strength of the spin-orbit interaction via an external electric field.   
The conditions under which the incoming and
transmitted spinors are connected unitarily will be determined, 
leading in principle, to a lossless single qubit gate. By connecting a few such
rings in an appropriate manner, one can achieve practically all the
important one qubit gates \cite{NC00}.

We consider a ring of radius $a$ in the $x-y$ plane and assume a tunable
static electric field \cite{NATE97} in the $z$ direction characterized by
the parameter $\alpha $. Then the spin dependent Hamiltonian \cite%
{Meijer,MPV} of a charged particle of effective mass $m^{\ast }$ is 
\begin{equation}
H=\hbar \Omega \left[ \left( -i\frac{\partial }{\partial \varphi }+\frac{%
\omega }{2\Omega }(\sigma _{x}\cos \varphi +\sigma _{y}\sin \varphi )\right)
^{2}-\frac{\omega ^{2}}{4\Omega ^{2}}\right] ,  \label{Ham}
\end{equation}%
where $\varphi $ is the azimuthal angle of a point on the ring,  
$\hbar\Omega =\hbar ^{2}/2m^{\ast }a^{2}$ is the parameter characterizing the
kinetic energy of the charge and $\omega$ =$\alpha /\hbar a$ is the
frequency associated with the spin-orbit interaction. Apart from constants,
the Hamiltonian (\ref{Ham}) is the square of the sum of the $z$ component of
the orbital angular momentum operator $L_{z}=-i\frac{\partial }{\partial
\varphi }$, and of $\frac{\omega }{\Omega }S_{r}$, where $S_{r}=\sigma
_{r}/2 $ is the radial component of the spin (both measured in units of $%
\hbar )$. $H$ commutes in a nontrivial way with $K=L_{z}+S_{z}$, the $z$ component 
of the total angular momentum.  $H$ also commutes with 
$S_{\theta\varphi }=
S_{x}\sin \theta \cos \varphi +S_{y}\sin \theta \sin \varphi
+S_{z}\cos \theta $, the spin component in the direction determined by the
angles $\theta $, and $\varphi $, where $\theta $ is given by 
\begin{equation}\tan \theta
=-\omega /\Omega . \label{tant}
\end{equation}
One easily can prove that the commutator 
$\left[ K,S_{\theta \varphi }\right] =0$ and therefore,
we may look for simultaneous eigenstates of $H$, $K$ and 
$S_{\theta \varphi}$. In the $\left\vert +\right\rangle$, $\left\vert -\right\rangle $ 
eigenbasis of $S_{z}$ one finds these in the form: 
\begin{equation}
\psi (\kappa ,\varphi )=e^{i\kappa \varphi }\binom{e^{-i\varphi /2}u(\kappa )%
}{e^{i\varphi /2}v(\kappa )}  \label{est}
\end{equation}%
obeying: 
\begin{subequations}
\label{KSev}
\begin{align}
K\psi (\kappa ,\varphi )& =\kappa \psi (\kappa ,\varphi )  \label{Kev} \\
S_{\theta \varphi }\psi (\kappa ,\varphi )& =s(\kappa )\psi (\kappa ,\varphi
),\quad s(\kappa )=\pm 1/2  \label{Sev}
\end{align}%
\end{subequations}
and the energy eigenvalues are 
\begin{equation}
E=\hbar \Omega \left[ \kappa ^{2}-\mu \kappa w+1/4\right] ,\quad \mu =\pm 1,
\label{En}
\end{equation}%
with $w=\sqrt{1+(\omega ^{2}/\Omega ^{2})}$. In a closed ring $\kappa \pm 1/2$
must be integer, while if one considers leads connected to the ring, there
is no such restriction: the energy is a continuous variable, and then the
possible values of $\kappa $ are the solutions of Eq. (\ref{En}), which can
be written as: 
\begin{equation}
\kappa _{j}^{\mu }=\mu (w/2+(-1)^{j}q),\qquad j=1,2,\quad \mu =\pm 1,
\end{equation}%
where $q=\sqrt{(\omega /2\Omega )^{2}+E/\hbar \Omega }$. The energy
eigenvalues are four fold degenerate: $j=1,2$ correspond to two distinct
values of  $\left\vert \kappa _{j}^{\mu }\right\vert $, while the additional
degeneracy at a given $j$ is resolved by the sign of $\kappa _{j}^{\mu }$.
The components of the eigenvectors in (\ref{est}) are related as: 
\begin{equation}
\frac{v(\kappa _{j}^{\mu })}{u(\kappa _{j}^{\mu })}=(\tan {\theta /2})_{\mu }%
=\frac{\Omega }{\omega }\left( 1-\mu w\right) .\text{\quad }  \label{theta}
\end{equation}
The two possible eigenvalues in  Eq. (\ref{Sev})  are  
$s(\kappa _{j}^{\mu})=-\mu /2$, accordingly  the eigenstates
for a given energy can be classified by giving the absolute
value of $\kappa _{j}^{\pm }$, together
with the sign of the eigenvalue of $S_{\theta \varphi }$.

The stationary states of the problem: ring plus leads, can be determined by
fitting the solutions obtained in the different domains. Using local
coordinates as shown in Fig.~1, the incoming wave, $\Psi _{I}(x)$, and the
outgoing wave $\Psi _{II}(x^{\prime })$\ are built up as linear combinations
of spinors with spatial dependence $e^{ikx}$ etc. corresponding to 
$E={\hbar ^{2}k^{2}}/{2m^{\ast }}$: 
\begin{equation}
\Psi _{I}(x)=\binom{f_{1}}{f_{2}}e^{ikx}+\binom{r_{1}}{r_{2}}e^{-ikx},\quad
\Psi _{II}(x^{\prime })=\binom{t_{1}}{t_{2}}e^{ikx^{\prime }}.
\end{equation}

\begin{figure}[tbh]
\begin{center}
\psfrag{in}[tl][tl]{$\Psi _{I}(x)$} \psfrag{out}{$\Psi _{II}(x^{\prime })$} %
\psfrag{gamma}{$\gamma$} \psfrag{up}[tl][tl]{$\Psi _{u}(\varphi)$} %
\psfrag{down}{$\Psi _{l}(\varphi^{\prime })$} \psfrag{p1}{$\varphi$} %
\psfrag{p2}{$\varphi^{\prime }$}
\par
\includegraphics*[width=6.5cm]{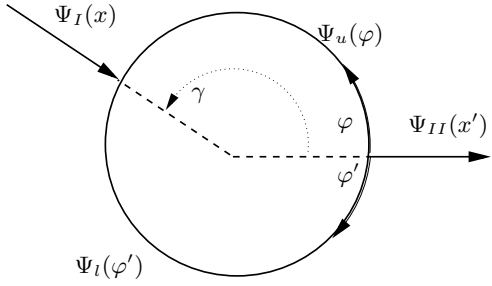}
\end{center}
\caption{ The geometry of the device and the relevant wave functions in the
different domains. }
\label{ringfig}
\end{figure}
The wave functions belonging to the same energy can be written as linear
combinations of the corresponding four eigenspinors in the upper and lower
arms of the ring as: 
\begin{equation}
\Psi _{u}(\varphi )=\sum_{\substack{ j=1,2  \\ \mu =\pm }}a_{j}^{\mu }\psi
(\kappa _{j}^{\mu },\varphi );\text{ \ }\Psi _{l}(\varphi ^{\prime })=\sum 
_{\substack{ j=1,2  \\ \mu =\pm }}b_{j}^{\mu }\psi (\kappa _{j}^{\mu
},-\varphi ^{\prime }),
\end{equation}%
respectively. According to Fig.~1 the incoming wave at $x=0$ is fitted to 
$\Psi _{u}$ at $\varphi =\gamma $ and to 
$\Psi _{l}$ at $\varphi ^{\prime}=2\pi -\gamma $, while the outgoing wave at
$x^{\prime }=0$ is fitted to
$\Psi _{u}$ and $\Psi _{l}$ at $\varphi =\varphi ^{\prime }=0$.
One has to require the continuity of the wave functions, as well as a vanishing spin
current density at the two junctions \cite{Griffith,Xia,MPV}. The resulting
set of linear equations leads to a relation between the expansion
coefficients in the different domains. The detailed procedure for the case 
$\gamma =\pi $ was described in Ref. \onlinecite{MPV}  using the eigenbasis
of $S_{\theta \varphi }$ at both junctions.
As we will show, the more general
geometry shown in Fig. 1. allows a significantly wider class
of spin transformations to be described now 
in the fixed $S_{z}$ basis, which is  more suitable to
discuss the qubit operations. 
We focus here on 
the transmission properties of the ring, and obtain in the
$\left\vert+\right\rangle$, $\left\vert -\right\rangle $ basis: 
\begin{equation}
\left( 
\begin{array}{c}
t_{1} \\ 
t_{2}%
\end{array}%
\right) =T\left( 
\begin{array}{c}
f_{1} \\ 
f_{2}%
\end{array}%
\right) =\left( 
\begin{array}{cc}
T_{11} & T_{12} \\ 
T_{21} & T_{22}%
\end{array}%
\right) \left( 
\begin{array}{c}
f_{1} \\ 
f_{2}%
\end{array}%
\right) ,
\end{equation}%
with 
\begin{equation}
T=\left\vert T_{\gamma }\right\vert e^{i\delta _{0}/2}e^{-i\gamma /2}U,
\label{Tmat}
\end{equation}%
where the matrix elements of $U$ are 
$u_{11}=u_{22}^{\ast }=
(e^{i\delta/2}\sin^{2}\frac{\theta }{2}
+e^{-i\delta /2}\cos ^{2}\frac{\theta }{2})e^{i\gamma /2}$,
$u_{12}=-u_{21}^{\ast }=i\sin \frac{\delta }{2}\sin\theta e^{-i\gamma /2}$.
$\left\vert T_{\gamma }\right\vert$ and the phases 
$\delta _{0}$ and $\delta $ are obtained from 
\begin{widetext}
\begin{eqnarray}
\left\vert T_{\gamma }\right\vert e^{i\delta _{\pm }}=\frac{4ikaq(\sin q(2\pi
-\gamma )+e^{i\Phi _{\pm }}\sin q\gamma )e^{-i\gamma \Phi _{\pm }/2\pi }}{%
k^{2}a^{2}\left\{ \cos 2q(\pi -\gamma )-\cos 2q\pi \right\} +4q^{2}\left\{
\cos \Phi -\cos 2q\pi \right\} +4ikaq\sin 2q\pi }, \\
\delta _{0}=\delta _{+}+\delta _{-},\qquad \delta =\delta _{+}-\delta
_{-}=2\arctan \frac{\sin w\gamma \sin q(2\pi -\gamma )+\sin w(2\pi -\gamma
)\sin q\gamma }{\cos w\gamma \sin q(2\pi -\gamma )-\cos w(2\pi -\gamma )\sin
q\gamma }, \label{del}
\end{eqnarray}%
\end{widetext}
where $\cos \Phi _{+}=\cos \Phi _{-}\equiv \cos \Phi $, with 
$\Phi _{\pm }=\pi (-1\pm w)$,
the Aharonov-Casher phases \cite{AC} for the corresponding spin directions.

The important fact is that $U$ is a unitary, unimodular matrix. It is this
unitary part that performs a nontrivial spin transformation in the qubit
space. In Eq. (\ref{Tmat}) $\left\vert T_{\gamma }\right\vert $ is a
non-negative constant with $\left\vert T_{\gamma }\right\vert \leq 1$, which
can be considered as the efficiency of the gate.
Therefore, one has in general 
$\left\vert t_{1}\right\vert ^{2}+\left\vert t_{2}\right\vert ^{2}\leq 1$,
nevertheless the transmitted amplitudes can be renormalized,
and their absolute value squared give the probabilities of having the
corresponding spin direction, if the particle is assumed to be transmitted
at all. In certain cases, to be discussed below, we find, however, that
$T$ is unitary: $|T_{\gamma }|=1$. We shall turn now to
analyze the transformation properties of this device in more detail.

If the incoming and outgoing leads are connected to the ring diametrically,
then $\gamma =\pi $, and as seen from (\ref{del}) also $\delta =\pi $
independently from the energy. The transmission matrix then takes the form :%
\begin{equation}
T=\left\vert T_{\pi }\right\vert e^{i(\delta _{0}+\pi )/2}\left( 
\begin{array}{cc}
\cos \theta & -\sin \theta \\ 
\sin \theta & \cos \theta%
\end{array}%
\right) =\left\vert T_{\pi }\right\vert e^{i(\delta _{0}+\pi) /2}U,
\label{TMpi}
\end{equation}%
with 
\begin{eqnarray}
|T_{\pi}|e^{i\delta _{0}/2} &=&   
\end{eqnarray} 
\begin{equation*}
\frac{8ikaq\sin (\pi q)\cos (\Phi /2)}
{k^{2}a^{2}(1-\cos 2q\pi)+4q^{2}(\cos\Phi -\cos 2q\pi )+4ikaq\sin 2q\pi}.
\end{equation*}
The unitary part $U$ of the transformation given by Eq. (\ref{TMpi}) is
independent of the wave vector $k$, and it rotates the spin around the $y$
axis \cite{NC00} by an angle $2\theta $. By changing the strength 
of the
spin-orbit interaction \cite{NATE97} $\omega =\alpha /\hbar a$, according
to Eq.(\ref{tant}) the values of $|\theta|$ can be varied from $0$ up to 
$0.8(\pi /2)$. Fig.~2 shows the gate efficiency 
$\left\vert T_{\pi}\right\vert $ as a function of $|\theta|$ and of
$ka$ around $k_Fa=20.4$, corresponding to a ring 
of radius $0.25\mu m$ and a Fermi energy $11.13$ meV of InGaAs. 
One sees also that for
several values of $ka$ and $\theta $ the transformation is strictly unitary,
with $\left\vert T_{\pi }\right\vert =1$. If one couples such unitary
devices in series, then obviously the resulting transformation will be the
product of the corresponding unitary matrices, and will be unitary again.

\begin{figure}[htb]
\begin{center}
\includegraphics*[width=8.3cm]{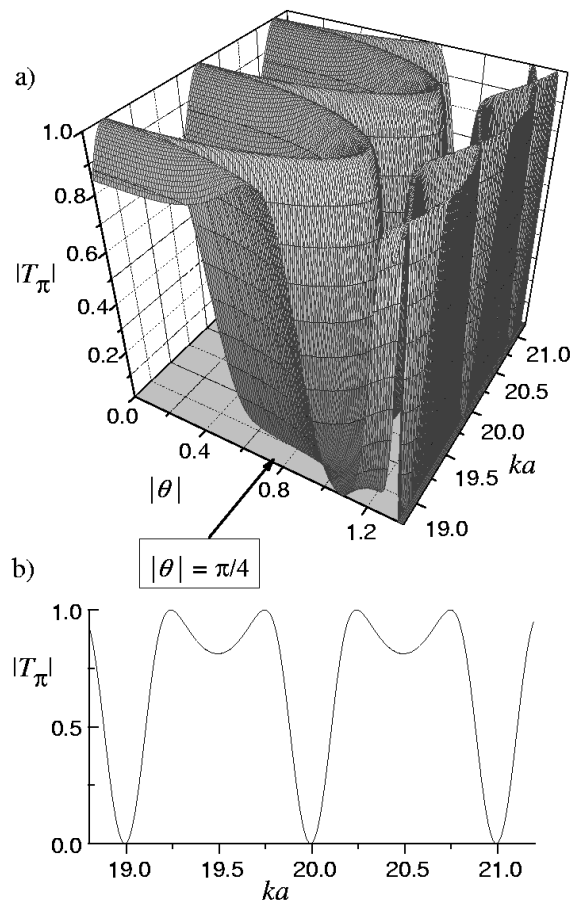}
\end{center}
\caption{(a) Efficiency $|T_{\protect\pi }|$ of the quantum gate with $\gamma=\pi$
as a function of the modulus of the half rotation angle
$\protect\theta=-\arctan{\protect\omega/\Omega}$ and $ka$. The maximal value
of $|\protect\theta|$ corresponds to $\protect\omega/\Omega=3.5$.
(b) Cross section of the surface at $|\protect\theta|=\protect\pi/4$,
where the transformation is essentially an Hadamard gate.}
\label{straight}
\end{figure}

In the language of quantum informatics \cite{NC00}, the transformation 
(\ref{TMpi}) above, represents a rather general single qubit gate, and it shows
that in principle a continuous set of spin rotations can be achieved with
already a single diametrically connected ring. This can be further extended
by coupling two or more such rings in series. A transformation of the form 
(\ref{TMpi}) with $\theta =\pi /4$ is essentially a so called Hadamard gate 
\cite{NC00}, which plays a distinguished role in quantum algorithms, while
two such gates in series results in an $X$ gate or quantum NOT
gate. Strictly speaking in both cases one has to introduce an additional
relative phase between the components, in order to have the correct
determinant $(-1)$ of the transformations \cite{NC00}. This is possible
with $\gamma \neq \pi$ to be discussed below.

In the case of arbitrary $\gamma $ other types of transformations can be
realized. An important particular case is when $\delta =0$, which can be
achieved by  tuning the  voltage, $\alpha $ and thereby 
$\omega /\Omega $. Then one has: 
\begin{equation}
T_{\gamma }(\delta =0)=|T_{\gamma }|e^{i\delta _{0}/2}\left( 
\begin{array}{cc}
1 & 0 \\ 
0 & e^{-i\gamma }%
\end{array}%
\right) ,  \label{phase}
\end{equation}%
and the unitary part of the transformation is a phase gate \cite{NC00},
where the phase difference introduced between spin up and spin down is just the
geometrical angle $\gamma $ (Fig.~1). 
Fig.~3 shows the curves along which 
such phase gates can be realized ($\delta=0$ in Eq.~(\protect\ref{del}))
depending on the values of $ka$ and 
$\omega/\Omega $. The dots on the curves mark the points where
$|T_{\gamma}|=1 $, and thus the transformation is unitary.
In Fig.~4 we show the gate efficiency of a phase gate
($\delta=0$) for the special value
$\gamma =\pi /2$ as function of $\alpha $ and $ka$.

\begin{figure}[htb]
\begin{center}
\includegraphics*[width=8.3cm]{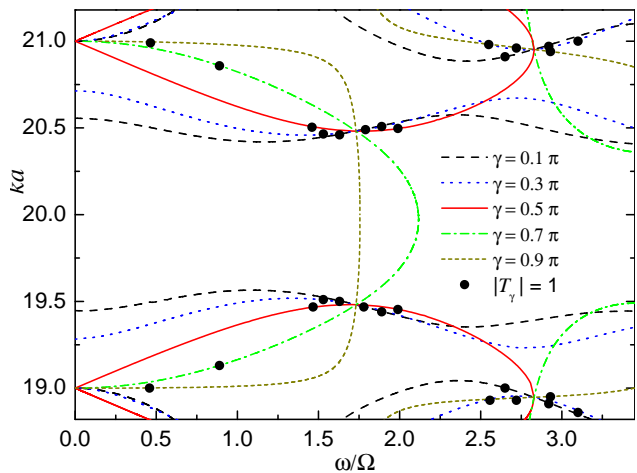}
\end{center}
\caption{ Lines along which the ring acts as a $\protect\gamma$ phase gate.
Points on the curves show where the gates are lossless i.e.  
$|T_{\protect\gamma}|=1$}
\label{phase1}
\end{figure}

\begin{figure}[htb]
\begin{center}
\includegraphics*[width=8.3cm]{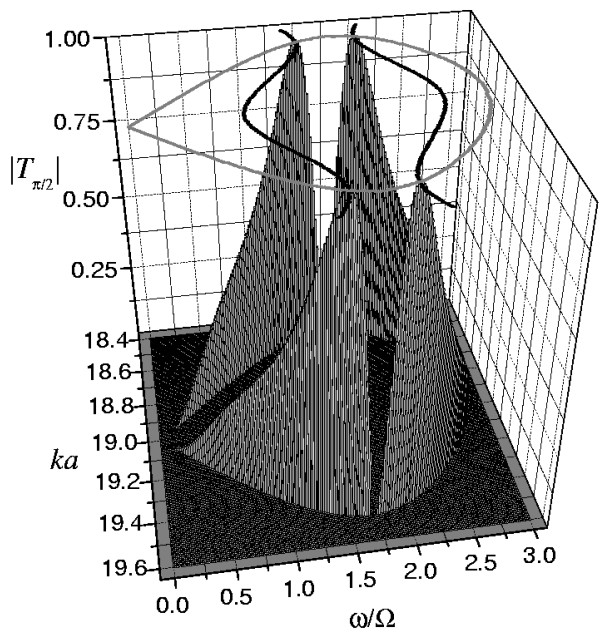}
\end{center}
\caption{Efficiency of a $\protect\pi/2$ phase gate. 
Along the gray curve on the top  
($ka$ -- $\protect\omega/\Omega$)
plane the ring acts as a $\protect\pi/2$ gate. The
black lines on the same plane show where the efficiency $|T_{\protect\pi/2}|$
equals unity, thus at the crossing points of the black and gray lines this
phase gate is unitary.}
\label{phase2}
\end{figure}

We note that in principle a number of other gates can be constructed by coupling
several of those rings. This can be realized with parameters corresponding to
unitary gates, so that the product of the corresponding spin rotations
result again in a unitary transformation. For instance, two rings both 
with $\gamma =\pi /2$ in Eq. (\ref{phase}) is a $Z$ gate \cite{NC00}. If such a
gate is coupled to a diametric ring associated to Eq.~(\ref{TMpi}) with $%
\theta =\pi /4$, one obtains exactly an Hadamard gate. Similarly, two rings
with $\gamma =\pi /2$, plus two of the type corresponding to 
Eq.~(\ref{TMpi}) with $\theta =\pi /4$, yields
a\ NOT gate with the correct phase.

In conclusion, we have shown that a quantum ring with Rashba type
spin-orbit interaction can serve as a one qubit quantum gate for electron
spins. 
The spin transformation properties of the gates can be extended by coupling 
such rings in series. Different types of gates can be realized 
by tuning the electric field strength and changing the
geometric position of the junctions connected to the ring, as well as by
fabricating rings with different sizes.
The considered parameters are within the experimentally feasible range 
\cite{NATE97,NMT99,SK01}.
In a ring of radius $a=0.25\mu m$
and for InGaAs ($m^{*}=0.023m$), $\alpha$
can be varied up to $2.0 \times 10^{-11}$
eVm \cite{NATE97}, which corresponds to $\theta=0.8(\pi/2)$.

We note that similar rings in the presence of an external magnetic field
can be used for spin filtering \cite{MPV}. This points to the
possibility to integrate gates and filters that can serve as elementary
building blocks of a quantum network based on spin sensitive devices 
\cite{EBL03,SB03,YPS02,FHR01}.

Acknowledgements: This work was supported by the Flemish-Hungarian Bilateral
Programme, by the Hungarian Scientific Research Fund (OTKA) under Contracts
Nos. D46043, M36803, and by the Hungarian Ministry of Education under
Contract No. FKFP 099/2001. One of the authors (B. M.) is supported by a
DWTC fellowship to promote S \& T collaboration between Central and Eastern
Europe.


\end{document}